# Generativism: Toward a Learning Theory for the Age of Generative Artificial Intelligence

Shan Li[1,2]
Email: shla22@lehigh.edu
ORCID: https://orcid/org/0000-0001-6001-1586

Juan Zheng[1]
Email: juz322@lehigh.edu
ORCID: https://orcid/org/0000-0002-4896-8198

[1]Department of Education and Human Services, College of Education, Lehigh University
[2]Department of Community and Global Health, College of Health, Lehigh University
Bethlehem, PA, USA, 18015

All correspondence should be addressed to Shan Li, HST Building 132, 124 E. Morton Street, Lehigh University, Bethlehem, PA 18015, USA, shla22@lehigh.edu

**Data availability statement**: No new data were generated or analyzed in this study.

**Conflict of Interest**: The authors declare that they have no conflict of interests.

**Abstract**

The four dominant learning theories of behaviorism, cognitivism, constructivism, and connectivism show significant conceptual limitations as generative artificial intelligence (AI) proliferates in educational settings. These frameworks were formulated before the emergence of AI systems capable of generating, synthesizing, and reasoning about knowledge. This article critically examines each learning theory and identifies assumptions challenged by generative AI's affordances. Drawing on research in distributed cognition, extended mind, human-AI collaboration, AI literacy, cognitive offloading, and metacognition, the article proposes *Generativism* as a learning theory for the generative AI age. Generativism posits that learning increasingly occurs through the iterative co-construction of knowledge between human learners and AI systems. The proposed framework is organized around four constructs (epistemic partnership, distributed agency, generative literacy, and adaptive metacognition) and four principles specifying the relations among them. The framework offers a foundation for rethinking instructional design, learning, assessment, and expertise development in contexts where generative AI plays an integral role in cognition.

# Generativism: Toward a Learning Theory for the Age of Generative Artificial Intelligence

Learning theories have long served as foundational frameworks for understanding how individuals acquire, process, and retain knowledge (Boghossian, 2006; Bransford et al., 2005; Ertmer & Newby, 2013). Four major learning theories have shaped educational research and practice: behaviorism (Watson, 1913, 2017), cognitivism (Garnham, 2019; Haugeland, 1978), constructivism (Amineh & Asl, 2015; Fosnot, 2013), and connectivism (Siemens, 2005b). Each theory has reflected the epistemological and technological conditions of its era. Behaviorism emerged during a period when psychology sought the objectivity of natural science (Watson, 1913). Cognitivism developed as computational metaphors of mind gained influence (Garnham, 2019; Neisser, 2014). Constructivism emerged amid growing recognition that learners actively construct meaning through experience and social interaction (Glassman, 2001; Piaget, 1968). Connectivism was formulated in response to the internet's transformation of how information is stored, accessed, and shared (Downes, 2022; Siemens, 2005a).

The emergence of generative artificial intelligence (AI) has introduced new challenges for educational theory. Unlike previous technologies, generative AI does not merely store or transmit information. It generates, synthesizes, and transforms knowledge artifacts in ways that can appear indistinguishable from human cognitive output (Giannakos et al., 2025; Yan et al., 2024). This capability raises questions that existing learning theories do not fully address: What happens when learners can produce sophisticated outputs before they fully understand the underlying concepts? How should learning be defined when cognitive processes can be distributed between human and machine? And when does the offloading of cognitive tasks to AI support learning, and when does it undermine it? These questions extend beyond practical concerns and raise important theoretical issues.

This article has two purposes. First, it provides a critical examination of the four major learning theories and evaluates how well each accounts for learning with generative AI. Second, drawing on recent scholarship in distributed cognition, extended mind, AI literacy, cognitive offloading, human-AI collaboration, and metacognitive research, the article introduces *Generativism* as a new theoretical framework. The use of *Generativism* in this article distinguishes itself from the term from Chomsky's (1965) generative grammar in linguistics. In this article, the term denotes a learning theory concerned with how humans learn in partnership with generative AI systems. The article contributes a set of constructs and principles that conceptualize this emerging form of learning. It offers a foundation for rethinking instructional design, learning, assessment, and expertise development in generative AI-rich environments.

## The Four Learning Theories: Foundations and Core Assumptions

### Behaviorism

Behaviorism emerged in the early twentieth century when Watson (1913) argued that the theoretical goal of psychology "is the prediction and control of behavior" (*p*. 158). Psychology should be treated as "a purely objective, experimental branch of natural science" (Watson, 1913, p. 176). The early work of Pavlov (1927) on classical conditioning showed that neutral stimuli can elicit reflexive responses after repeated pairing. Building on this foundation, Watson (2017) held that learning is the process of forming associations between environmental stimuli and behavioral responses. Skinner (1963) extended the behaviorist framework with his theory of operant conditioning, which emphasized the role of consequences, such as reinforcement and punishment, in shaping behavior. Skinner's (2014) research demonstrated that organisms tend to repeat behaviors that produce favorable outcomes and suppress those that produce unfavorable ones. Furthermore, Bandura's (1963) social learning theory advanced behaviorist principles by revealing that learning can occur through observation and modeling rather than solely through direct reinforcement. The influence of behaviorism on educational practice has been substantial and enduring. From a behaviorist perspective, instruction should emphasize clear objectives, incremental progression, and immediate feedback. Programmed instruction, computer-based drill-and-practice software, standardized testing, and behavior management systems all trace their conceptual lineage to behaviorist principles (Burton et al., 1996).

Several core assumptions underlie the behaviorism theory. First, learning is defined as a change in observable behavior, and internal mental processes are either ignored or treated as beyond the scope of scientific inquiry. Second, the learner is conceptualized as relatively passive, shaped by environmental contingencies rather than by internal cognitive processes. Third, knowledge is viewed as objective and external to the learner, transmitted and reinforced through stimulus-response pairings. Fourth, the instructor’s role is to structure the environment so that desired behaviors are reinforced (Ertmer & Newby, 2013). Despite widespread critiques (Harzem, 2004; Mapel, 1977; Schnaitter, 1999), behaviorist approaches continue to inform instructional practice in areas such as skill acquisition, behavior modification, and competency-based education.

**Cognitivism**

Cognitivism arose in the 1950s and 1960s as a reaction to behaviorism’s limitations in explaining complex mental phenomena such as language acquisition, problem-solving, and creative thought (Chomsky, 1965; Neisser, 2014). Influenced by advances in computer science, neuroscience, and information theory, cognitivist theorists proposed that the mind functions as an information processor that receives, encodes, stores, and retrieves information through structured mental operations (Atkinson & Shiffrin, 1968; Miller, 1956). Piaget and Cook's (1952) theory of cognitive development was instrumental in establishing the view that learners actively construct mental schemas through processes of assimilation and accommodation. Piaget (1970) argued that cognitive development proceeds through distinct stages, each characterized by increasingly sophisticated modes of thought.

Ausubel's (1968) theory of meaningful learning further advanced cognitivist thinking by emphasizing the role of prior knowledge in acquiring new information. Ausubel (1968) proposed that learning is most effective when new material is connected to existing cognitive structures through “advance organizers” that scaffold the integration of unfamiliar concepts. Bruner's (1961) theory of discovery learning similarly stressed the importance of active cognitive engagement, proposing that learners construct understanding most effectively when they discover principles and relationships for themselves. Information processing theory, drawing on the work of Atkinson and Shiffrin (1968) and later elaborated by Sweller's (1988) cognitive load theory, described how the limitations of working memory constrain learning and how instructional design can manage intrinsic, extraneous, and germane cognitive load. These models had direct implications for instructional design. They indicated that prior knowledge should be activated to facilitate encoding, information should be presented in manageable chunks, and practice should be spaced and varied to promote transfer.

The central assumptions of cognitivism include the following: (1) learning involves internal mental processing, (2) the mind operates through structured schemas and mental models, (3) prior knowledge influences the acquisition of new knowledge, (4) instructional design should align with how the brain processes information, and (5) metacognitive strategies can enhance learning effectiveness (Ertmer & Newby, 2013; Garnham, 2019; Neisser, 2014). Cognitivism’s emphasis on the internal architecture of cognition represented a significant advance over behaviorism's exclusive focus on observable behavior.

**Constructivism**

Constructivism represents a broad family of theories united by the epistemological claim that knowledge is not passively received but actively constructed by the learner (Fosnot, 2013; Fosnot & Perry, 1996). Two major strands dominate the constructivist landscape (Amineh & Asl, 2015). Cognitive constructivism, rooted in the work of Piaget (1968, 1970), holds that individuals build knowledge through direct interaction with their environment. It proposes that learners construct and revise mental models in response to new experiences. Social constructivism, developed primarily by Vygotsky (1978), emphasizes the role of social interaction, cultural tools, and language in the construction of knowledge.

Vygotsky's concept of the Zone of Proximal Development (ZPD) has been particularly influential in educational theory. The ZPD describes the distance between what a learner can accomplish independently and what can be achieved with guidance from a more knowledgeable other (Vygotsky, 1978). The related concept of scaffolding, elaborated by Wood et al. (1976), describes the process by which an expert provides temporary support that is gradually withdrawn as the learner develops competence. Social constructivism also draws on the work of situated cognition theorists such as Brown et al. (1989) and Wilson (1993), who argued that knowledge is embedded in the contexts and activities in which it is used. Von Glasersfeld's (1984) radical constructivism pushed the epistemological implications further, arguing that knowledge does not represent an objective reality but is instead a viable construction that fits the learner's experience. Lave and Wenger's (1991) theory of situated learning and communities of practice added another dimension by arguing that learning is embedded in social contexts and that becoming a competent member of a community is itself a form of learning.

Constructivism rests on several foundational assumptions: (1) knowledge is constructed, not transmitted; (2) learning is an active process requiring engagement and effort; (3) social interaction plays a central role in knowledge construction; (4) prior knowledge and cultural context shape how new information is interpreted; and (5) authentic, contextualized tasks promote deeper learning than decontextualized instruction (Duffy & Cunningham, 1996; Ertmer & Newby, 2013; Fosnot, 2013). Constructivism has had a profound influence on pedagogical practice, informing approaches such as problem-based learning, inquiry-based instruction, and collaborative learning (Ertmer & Newby, 2013).

**Connectivism**

Connectivism was proposed as a learning theory for the digital age (Downes, 2022; Siemens, 2005b). Siemens (2005a) argued that existing theories of behaviorism, cognitivism, and constructivism were developed before the advent of digital technologies and therefore fail to account for how learning occurs in networked, information-rich environments. Siemens (2005b) asserted that these theories "do not address learning that occurs outside of people (i.e., learning that is stored and manipulated by technology)" *(p*. 4). According to Siemens (2005a), learning is the process of creating connections between nodes of information distributed across networks, including human and non-human nodes. Downes (2022) further articulated connectivism as a framework in which "knowledge is distributed across a network of connections, and therefore that learning consists of the ability to construct and traverse those networks" (*p* . 59). However, connectivism's status as a distinct learning theory has been debated. Verhagen (2006) argued that connectivism describes a pedagogical view rather than a learning theory. Kop and Hill (2008) noted that the theory's claims overlap substantially with constructivism theory. Despite these critiques, connectivism provided the conceptual foundation for massive open online courses (MOOCs) and other forms of networked learning (Clarà & Barberà, 2013).

Based on Siemens' (2005a) eight principles, connectivism is grounded in the following assumptions: (1) knowledge emerges from the diversity of perspectives within a network; (2) learning is a process of forming connections across distributed nodes and information sources; (3) knowledge can exist outside of individual mind; (4) the capacity to acquire new knowledge matters more than what is currently known; (5) connections must be maintained to support ongoing learning; (6) recognizing patterns and relationships across disciplines is a key competency; (7) currency and accuracy of knowledge is central to all learning activities; and (8) decision-making is itself a learning process, as what counts as valid knowledge may shift with changes in the information landscape.

## What Generative AI Reveals About the Limits of Existing Theories

The emergence of generative AI does not simply add a new tool to the learning environment; it introduces a different kind of cognitive partnership. The following analysis identifies specific assumptions in each theory that are challenged by this development.

### Behaviorism and the Problem of Indistinguishable Output

Behaviorism defines learning as a change in observable behavior (Boghossian, 2006; Burton et al., 1996). If a student can produce a correct essay, solve a problem, or generate a functional code snippet, behaviorist theory would identify this as evidence of learning. Generative AI challenges this reasoning. A learner using a large language model (LLM) can generate essays, analyses, and code that appear indistinguishable from the work of someone who has developed the relevant understanding or competence (Yan et al., 2024). The observable behavior is identical, but the learning may not have occurred. When external outputs can be produced by delegating cognitive work to an AI system, observable behavior alone is insufficient to determine whether learning has taken place.

The dissociation between performance and learning is not a critique of behaviorism unique to the generative AI age. As pointed out by Soderstrom and Bjork (2015), current performance is an unreliable index of durable learning, because conditions that elevate immediate performance often fail to produce lasting change. Nevertheless, generative AI can widen the gap between performance and learning. For instance, Bastani et al. (2025) conducted a randomized field experiment with nearly one thousand high school students in mathematics. They found that students with access to a standard GPT-4 interface performed 48 percent better on practice problems. On a subsequent unassisted exam, however, they scored 17 percent worse than students who had practiced without AI. A tutored version of the system with pedagogical guardrails removed the negative effect but produced no lasting advantage.

**Cognitivism and the Redistribution of Cognitive Labor**

Cognitivist theory focuses on internal mental processes and on how information is encoded, stored, organized, and retrieved (Atkinson & Shiffrin, 1968; Garnham, 2019; Haugeland, 1978). Cognitive load theory, for example, addresses how instructional design can manage the demands placed on working memory (Sweller, 1988). However, with generative AI, learners can now externalize not merely routine information retrieval but also higher-order cognitive operations such as synthesis, argumentation, analysis, and creative production (Borge et al., 2024; Zhai et al., 2025). When an AI system synthesizes information, identifies patterns, and produces organized outputs, it performs functions that, under cognitivist theory, would constitute evidence of learning if performed by a human.

Moreover, recent research has identified a pattern of *cognitive offloading* in which learners delegate cognitive tasks to AI systems, thereby reducing their own cognitive engagement (Fan et al., 2025; Gerlich, 2025; Risko & Gilbert, 2016). Fan et al. (2025) found that learners using generative AI demonstrated reduced engagement in self-regulated learning processes such as reflection and self-evaluation, a pattern they termed “metacognitive laziness.” Gerlich (2025) reported a significant negative correlation between frequent use of AI tools and critical thinking abilities, with cognitive offloading as a mediating factor. The redistribution of cognitive work points to a conceptual limitation within cognitivism, as it does not fully account for learning processes that unfold across both human and AI systems.

The challenge to cognitivism extends beyond the argumentation that cognitive work can be shared. Cognitive load theory holds that schema acquisition depends on effortful processing in working memory, and instructional design succeeds when it directs this limited capacity toward learning-relevant processing (Sweller et al., 1998). When a LLM performs the selection, organization, and integration of information, learners have fewer opportunities to engage in the cognitive work through which schemas are built. Sweller et al. (1998) noted that humans acquire most knowledge by borrowing it from others, yet borrowing still requires effortful reorganization within the learner’s cognitive architecture. AI output is usually already organized, which encourages learners to retain it without reorganizing it themselves. For instance, Stadler et al. (2024) compared university students who used a LLM with students who used conventional web search during a scientific inquiry task. Students in the AI condition reported significantly lower mental effort, yet they produced reasoning of significantly lower quality. A related problem concerns the accuracy of learners’ self-knowledge. People often mistake the ability to access external information for evidence of their own internal understanding or competence. Fisher et al.

(2015) showed that searching the internet inflates people's estimates of their own knowledge. Sparrow et al. (2011) found that people who expect information to remain available tend to remember where it is stored rather than what it says. Rozenblit and Keil (2002) demonstrated that people hold a persistent illusion of explanatory depth even in low-technology conditions. Generative AI may increase these risks, because its output is more fluent, personalized, and complete in form.

**Constructivism and the Reduction of Effortful Meaning-Making**

Constructivism holds that learners actively build knowledge through experience, social interaction, and reflection on prior understanding (Glassman, 2001; Piaget, 1970; Vygotsky, 1978). According to constructivism theory, each learner's prior knowledge and lived experience serve as the foundation for new understanding (Piaget, 1970). For constructivists, learning depends not simply on exposure to information but on the learner's active transformation of that information into personally meaningful knowledge. Chi and Wylie's (2014) ICAP framework predicts that learning improves as engagement moves from passive through active and constructive to interactive modes. Generative learning theory makes a similar prediction. Meaningful learning depends on selecting relevant information, organizing it into coherent structures, and integrating it with prior knowledge (Fiorella & Mayer, 2016; Wittrock, 1974). However, studies indicate that students often adopt AI-generated solutions quickly, bypassing the effortful meaning-making that constructivist learning requires (Zhai et al., 2024).

Constructivism also emphasizes learning as a social and collaborative process, where dialogue and negotiation among peers drive the co-construction of knowledge (Vygotsky, 1978). Generative AI can weaken this collaborative dynamic by providing responses that reduce the need for peer interaction and collective inquiry. These challenges suggest that educators working within a constructivist paradigm must critically reconsider how active knowledge construction is defined and facilitated in AI-mediated learning environments.

**Connectivism and the Shift from Navigation to Co-Generation**

Connectivism claims that learning occurs through forming connections across distributed nodes of information and various information sources (Downes, 2022; Siemens, 2005b). Knowledge can reside in non-human appliances such as databases, search engines, and information repositories (Siemens, 2005b). However, connectivism was formulated in an era when technology primarily stored and transmitted existing information. The “non-human appliances” contain knowledge but do not generate new knowledge in response to novel queries. In contrast, Generative AI generates novel text, produces original analyses, and synthesizes information in ways responsive to the specific framing of a question (Giannakos et al., 2025).

Moreover, connectivism assumes that learners can move through a visible network of information sources. In such a network, learners can see where information comes from, compare different sources, and judge which ones are credible. However, research on online evaluation shows that even this kind of source judgment is difficult. Breakstone et al. (2021) found that most high school students in a national sample could not reliably distinguish credible online content from untrustworthy content. Wineburg and McGrew (2019) similarly showed that even highly educated adults often evaluated unfamiliar websites poorly unless they used the lateral reading strategies practiced by professional fact-checkers. Given that people already struggle to evaluate information when sources are visible, the challenge becomes even greater when LLMs provide answers without making their sources clear. The primary challenge for learners is no longer navigating a network to find existing information (the “know-where” emphasized by connectivism) but evaluating, guiding, and integrating the AI-generated output.

Learners must now develop skills to assess the quality, accuracy, and relevance of the output and to iteratively refine prompts and outputs through sustained dialogue.

Table 1 summarizes each learning theory's core assumptions alongside the challenges introduced by generative AI.

**Table 1**

*Summary of Core Assumptions and Generative AI Challenges for Each Learning Theory*

| Learning Theory | Core Assumption | Nature of the Challenge |
|---|---|---|
| Behaviorism | Learning = observable behavior change | AI enables the production of sophisticated behavioral outputs without underlying cognitive change (behavioral equivalence problem) |
| Cognitivism | Learning = internal information processing | AI performs cognitive operations (synthesis, inference) previously exclusive to human minds; cognitive offloading redistributes cognitive labor |
| Constructivism | Learning = active construction of meaning through experience | AI provides ready-made knowledge that bypasses effortful meaning-making; learners consume rather than construct understanding |
| Connectivism | Learning = forming and navigating networked connections | AI subsumes information sources and networks; evaluating AI-generated output becomes more critical than navigating distributed knowledge |

## Toward Generativism: Theoretical Foundations and Synthesis

The critique above identifies a set of learning phenomena that existing theories do not fully address. The following section draws on four bodies of scholarship to construct the theoretical foundations of Generativism. These include (a) distributed cognition and extended mind theory, (b) human-AI collaboration and hybrid intelligence, (c) generative AI literacy, and (d) research on cognitive offloading and metacognitive engagement.

### Distributed Cognition and the Extended Mind Theory

The theory of distributed cognition provides a foundation for understanding learning in human-AI systems (Hollan et al., 2000; Hutchins, 1995). Hutchins (1995) argued that cognitive processes are not confined to individual minds but are distributed across individuals, artifacts, and environments. Researchers have used the distributed cognition framework to analyze how cognitive processes are shared between human operators and AI systems in domains such as aviation, remote operations, and collaborative problem-solving (Plebe et al., 2022). The extended mind theory further argued that cognitive processes can constitutively include elements of the external environment (Clark & Chalmers, 1998). According to Clark and Chalmers (1998), an external resource can be considered part of the cognitive system if it plays the same functional role as an internal cognitive process and is reliably available, automatically endorsed, and readily accessible. The philosophical positions on distributed cognition and the extended mind have significant implications for understanding learning with generative AI, or generativism. Learning is not something happening *in* the learner while they *use* AI; it is something emerging *across* the interaction. In the same vein, knowledge is not transmitted or individually constructed but co-constructed through human-AI interaction. The learner brings intentionality, lived experience, and critical judgment. The AI provides pattern recognition across vast corpora and generative capacity. Generativism frames AI from “tool” to “cognitive partner” (Cain, 2025; Creely & Blannin, 2025).

## Human-AI Collaboration and Hybrid Intelligence

Recent developments in human-AI collaboration offer a strong foundation for understanding how learners and AI can co-generate knowledge. A key contribution comes from research on human-algorithm centaurs, which refer to hybrid models that blend formal analytics with human intuition through symbiotic learning (Saghafian & Idan, 2024). Unlike conventional human-in-the-loop approaches, where the human serves primarily as a data annotator, centaur models treat the human as an equal partner whose subjective input such as preferences, decisions, and judgments is directly incorporated into the learning process (Saghafian & Idan, 2024). The centaur models of human-AI collaboration reframe the learner not as a passive recipient of AI-generated outputs but as an active contributor whose cognitive engagement shapes the system's behavior. The centaur model has demonstrated that combining human expertise with algorithmic reasoning can outperform either humans or AI alone, even in domains where human intuition is relatively weak (Saghafian & Idan, 2024). For instance, Orfanoudaki et al. (2022) proposed a human-algorithm centaur model for clinical care by "systematically enhancing algorithmic performance with human-based intuition" (*p*. 1). They found that the model outperformed human experts and the best machine learning algorithm.

Additionally, Fragiadakis et al. (2024) proposed a structured framework for evaluating human-AI collaboration (HAIC) that identifies three distinct modes: AI-centric, human-centric, and symbiotic. The symbiotic mode describes a balanced partnership characterized by two-way interaction, shared decision-making, and a continuous exchange of feedback aimed at achieving collective goals. In the symbiotic mode, both human intuition and AI's capabilities are essential for optimal outcomes (Fragiadakis et al., 2024). Similarly, Molenaar (2022) introduced the concept of hybrid human-AI intelligence as emerging learning technologies. Drawing on the augmentation perspective, hybrid intelligent systems are developed to leverage human strengths, compensate for human weaknesses, and behave more intelligently together than either could alone (Akata et al., 2020; Molenaar, 2022). In the educational context, Molenaar (2022) argued that AI and human cognition in learning are intertwined and entangled, and therefore, the system must be viewed and researched as a hybrid human-AI system. Her six levels of automation model further articulates how the transition of control between teacher, learner, and AI can be calibrated to support meaningful learning at different levels of complexity.

In sum, these frameworks carry important implications for a generativist view of learning. First, the quality of learning depends on the nature and depth of the partnership between the learner and the AI system. Second, effective human-AI collaboration requires a distinctive set of competencies, including knowing when to rely on AI output, when to override it, and how to structure the interaction to maximize joint performance. Symbiotic learning, in particular, implies that the learner's intuitive and metacognitive contributions are necessary for the system to function effectively (Saghafian & Idan, 2024). Third, the aim of education in the AI age should cultivate learners who can collaborate effectively with AI to achieve outcomes that neither could achieve alone. The augmentation perspective reinforces this by positioning AI not as a replacement for human thought but as a means to extend and enrich the learner's generative capacity (Molenaar, 2022).

**Generative AI Literacy**

The concept of AI literacy has emerged as a focal area of research, though its definition remains contested (Allen & Kendeou, 2024; Long & Magerko, 2020; Ng et al., 2021). Allen and Kendeou (2024) proposed the ED-AI Lit framework for education, organized around six components of *Knowledge*, *Evaluation*, *Collaboration*, *Contextualization*, *Autonomy*, and *Ethics*. This framework stresses the importance of developing a deep understanding of how AI systems function and critically evaluating their implications for learning and society. Long and Magerko (2020) synthesized 17 AI literacy competencies through an interdisciplinary scoping review of research on AI education and public understanding of AI. The competencies address recognizing AI systems, understanding how they work, interpreting data and model behavior, and reasoning about ethical and societal implications of AI use. The introduction of generative AI has prompted calls for a more specific form of literacy. Several scholars have argued that existing AI literacy frameworks do not adequately address the unique challenges posed by generative systems (Annapureddy et al., 2025; Bozkurt, 2024). Generative systems do not simply retrieve or classify information, but actively produce knowledge artifacts that learners must evaluate, refine, and integrate. Therefore, generative AI literacy encompasses not only technical understanding and ethical awareness but also the capacity to evaluate, refine, and integrate AI-generated outputs in support of genuine understanding and learning (Annapureddy et al., 2025).

Generative AI literacy provides an important conceptual foundation for the proposed generativism learning theory. Generative AI literacy underscores the central role of learners' interpretive, metacognitive, and epistemic activities when co-working with AI systems for knowledge production (Annapureddy et al., 2025). These capacities move beyond functional or ethical competence with AI tools. Generativism theorizes how such literate practices turn into a learning process defined by purposeful thinking, continual monitoring of understanding, and deliberate decision-making about when and how to rely on AI.

**Cognitive Offloading, Metacognition, and the Paradox of AI-Assisted Learning**

Research on cognitive offloading provides critical evidence for understanding the risks and opportunities of learning with generative AI. Risko and Gilbert (2016) defined cognitive offloading as the use of external aids to reduce the internal cognitive demands of a task. While cognitive offloading can be adaptive by freeing mental resources for higher-order processing, recent studies suggest that excessive offloading to AI systems may undermine the cognitive engagement necessary for deep learning. For instance, Fan et al. (2025) found that learners using generative AI showed reduced engagement in key self-regulated learning processes, despite short-term performance improvements. Yan et al. (2024) argued that generative AI presents both promises and challenges for human learning, depending on how the technology is integrated into educational practice. Effective learning with generative AI requires adaptive metacognition (Dunlosky & Metcalfe, 2008; Winne & Azevedo, 2014), which is the capacity to monitor and regulate the allocation of cognitive effort between self and AI to support genuine understanding. Tomisu et al. (2025) proposed the "Cognitive Mirror" paradigm, a framework for AI-powered metacognition and self-regulated learning. The paradigm positions AI as a metacognitive partner that prompts learners to articulate, inspect, and refine their own thinking. Tomisu et al. (2025) contended that any learning theory adequate to the generative AI age must foreground metacognition as a central construct. When AI can perform portions of the cognitive work, what differentiates learning that is enhanced by AI from learning that is replaced by AI is the learner's capacity to monitor understanding, detect over-reliance, and deliberately reallocate effort. For Generativism, this means the learner must develop metacognitive awareness of when to rely on AI-generated output and when to override it.

## Generativism: A Framework for Learning in the Age of Generative AI

Drawing on the theoretical foundations outlined above, this section presents Generativism as a framework for learning in the generative AI age. Generativism positions learning as neither purely human nor AI-driven but as an emergent process at the intersection. Generativism defines learning as the deliberate co-construction of knowledge through iterative human-AI interaction. Generativism can be articulated through four core constructs. These constructs serve as conceptual anchors that identify the distinctive features of learning in the age of generative AI. Four principles then specify the relations among the constructs (Figure 1).

### Epistemic Partnership

Generativism posits that learning in the generative AI age increasingly occurs through epistemic partnerships between human learners and AI systems. Wu et al. (2025) argue that human-GenAI learning should be understood as shared epistemic agency. The central educational risk is a loss of human epistemic agency when learners accept AI output without questioning, verification, or sense-making. Their framework further clarifies that learners approach AI with different epistemic stances (absolutist, multiplist, evaluativist), which shape whether AI functions as a shortcut to answers or as a genuine partner in inquiry.

The concept of epistemic partnership extends distributed cognition (Hollan et al., 2000) and the centaur model (Orfanoudaki et al., 2022; Saghafian & Idan, 2024). In Hutchins' (1995) framework, the artifacts involved in distributed cognition (e.g., charts, instruments, logs) typically function as external representations that store, display, or transform information. By contrast, generative AI operates as an active partner in the cognitive system. Generative AI produces explanations and solutions and adapts its responses to the evolving context of the interaction. This shift from artifact to interactive partner marks a crucial change in how “non-human” elements participate in cognition, and it motivates treating human-AI work as a partnership rather than an instance of tool-mediated distribution alone.

Epistemic partnership reframes the relationship between expertise and AI use. According to research on human-algorithm centaurs (Saghafian & Idan, 2024), the most effective teams were not those with the strongest individual components, but those with the best processes for integrating human and machine contributions. Similarly, the goal of education under Generativism is not to develop learners who can replicate what AI does but to develop learners who can form effective partnerships with AI. In the generative AI age, the development of expertise increasingly includes the metacognitive capacity to manage an epistemic partnership with AI, rather than merely the possession of domain knowledge.

Empirical research on human-AI teaming reinforces the claim that partnership quality determines outcomes. Vaccaro et al. (2024) meta-analyzed 106 experimental studies comparing humans, AI systems, and human-AI combinations. They found that, "on average, human-AI combinations performed significantly worse than the best of humans or AI alone" (*p*. 2293). This finding challenges the assumption that adding AI to a task automatically improves performance. Moreover, Vaccaro et al. (2024) found that human-AI combinations produced performance losses in decision-making tasks but showed stronger gains in content creation tasks, where humans and AI systems could iteratively generate, evaluate, and refine artifacts. These findings suggest that the value of human-AI work depends less on access to AI itself than on the quality of the interaction. A productive epistemic partnership requires learners A productive epistemic partnership requires learners to question, verify, interpret, and build on AI-generated contributions (Chinn et al., 2011; Wu et al., 2025).

**Distributed Agency**

In traditional learning theories, the learner is often treated as the primary agent of learning: the one who processes, constructs, or navigates knowledge. Generativism underscores distributed agency, which describes the allocation of cognitive work between human learners and AI systems. Distributed agency requires learners to develop awareness of when they are offloading cognitive tasks to AI, to make deliberate decisions about which tasks to delegate and which to retain, and to monitor the consequences of this distribution for their own learning. Kong (2026) distinguishes Decision-, Action-, and Outcome-level agency, arguing that people can experience agency not only in acting and producing outcomes but also in forming intentions and commitments. In AI-mediated learning, this upstream *Decision-level Agency* can be weakened when the system steers which options feel available or which plans seem “best,” leading learners to adopt intentions they have not fully authored. Distributed agency therefore requires attention not only to what learners do with AI, but also to how their intentions, decisions, and commitments are shaped through interaction with AI systems (Kong, 2026).

Agency in generativism is not transferred from the learner to the AI system, nor is it located exclusively in either one. Agency is enacted through negotiation: the learner guides the system, the system responds with probabilistic outputs, and meaning is constructed through cycles of interaction. In this process, the learner may set goals, frame problems, pose prompts, evaluate responses, revise directions, and decide when an output is useful or inadequate. At the same time, the AI system shapes the learner’s activity by making certain possibilities more visible, offering particular formulations, and introducing directions the learner may not have anticipated. Therefore, distributed agency refers to an interactive configuration in which human intention and machine generation continually influence one another (Wu et al., 2025). Learning occurs as learner engages in iterative cycles of prompting, interpreting, questioning, and revising.

Distributed agency also raises questions of epistemic responsibility (Lloyd, 2025). When a learner and an AI system jointly produce a knowledge artifact, authorship may be distributed, but responsibility cannot be fully delegated to the system. Generativism holds that the human learner retains ultimate responsibility for the accuracy, appropriateness, and ethical use of AI-supported work. This position is consistent with Wu's et al. (2025) argument that content produced through human-AI collaboration should ultimately remain the responsibility of the human partner. AI participation does not dissolve human accountability.

**Generative Literacy**

Recent scholarship has begun to specify what it means to be literate in the use of generative AI (Annapureddy et al., 2025; Bozkurt, 2024). For example, Annapureddy et al. (2025) proposed twelve defining competencies of generative AI literacy, offering a broad taxonomy of the knowledge and skills learners may need when working with generative systems. Building on this contribution, Generativism defines *generative literacy* as a form of critical literacy oriented toward prompting, evaluating, refining, and integrating AI-generated knowledge artifacts. To keep the construct theoretically aligned with learning as iterative co-construction, Generativism synthesizes four higher-level capacities that are most directly tied to learning through human-AI interaction. First, *prompt competence* is the ability to formulate and revise queries in ways that elicit useful responses. Second, *evaluative judgment* is the ability to assess AI-generated output for accuracy, coherence, relevance, and contextual appropriateness. Third, *epistemic vigilance* is the disposition to maintain critical scrutiny of AI-generated content even when it appears fluent, confident, or plausible. Fourth, *integrative synthesis* is the ability to combine AI-generated content with one's own prior knowledge, disciplinary understanding, and learning goals to produce meaning that exceeds the AI output itself.

Under Generativism, generative literacy is not merely a technical skill or an optional supplement to learning. It is an important condition for meaningful learning in generative AI environments. Learners need generative literacy in order to transform AI output into understanding, rather than simply consuming it as a replacement for their own thinking. This requires them to question, revise, contextualize, and integrate generated content through active engagement. In this sense, generative literacy supports the broader aims of Generativism by enabling learners to participate productively in AI-mediated learning while still developing their own understanding, judgment, and capacity for independent thought.

**Adaptive Metacognition**

Metacognition, or thinking about one's own thinking, has been recognized as central to effective learning (Borkowski, 1996; Kaplan, 2008; Winne & Azevedo, 2014). The research on metacognitive laziness (Fan et al., 2025), cognitive offloading (Gerlich, 2025; Risko & Gilbert, 2016), and the cognitive paradox of AI-assisted learning (Yan et al., 2024) all point to the centrality of metacognitive regulation as a determinant of whether AI enhances or undermines learning. Adaptive metacognition, as conceptualized within Generativism, refers to the learner's ongoing, feedback-driven regulation of learning in a setting where part of the cognitive work is performed externally by AI. Learners continually recalibrate how they use AI based on cues about comprehension, error, and progress.

The concept of adaptive metacognition builds on but extends existing models of self-regulated learning (Li & Lajoie, 2021; Winne & Hadwin, 1998; Zimmerman, 2000). Zimmerman's (2000) social cognitive model describes self-regulation as a cyclical process of forethought, performance, and self-reflection. Winne and Hadwin's (1998) model emphasizes the role of monitoring and control in regulating cognitive processes during learning. These models provide valuable foundations, but they were designed for contexts in which the primary regulatory challenge is managing one's own cognitive processes. In the generative AI context, learners must manage not only their own cognition but also the contribution of an AI system that is itself performing cognitive operations. These expanded regulatory demands are important because learners may not accurately judge what they understand when AI makes information feel easy to process. Bjork et al. (2013) argued that learners often use fluency and ease of processing as cues for judging learning, which can lead them to overestimate their understanding when material feels easy. Because AI-generated text is highly fluent, it creates favorable conditions for such illusions of competence. Adaptive metacognition therefore requires learners to question whether ease of comprehension reflects genuine understanding.

Adaptive metacognition involves learners actively monitoring and regulating how they engage with generative AI before, during, and after the learning process. Before using AI, learners assess the task, identify what they already know, and decide whether AI support is appropriate and which parts of the task can be delegated to AI. During interaction, they evaluate AI-generated outputs, detect gaps or inaccuracies, revise prompts, and adjust their strategy based on whether the AI response supports their learning goals. After the interaction, they reflect on how AI shaped their thinking, what they learned from the interaction, and whether the distribution of cognitive work strengthened or weakened their own understanding. Järvelä et al. (2023) further suggest that AI systems can participate in regulation by helping learners monitor progress and sustain reflection; however, such support is valuable only when it strengthens the learner's own regulatory activity.

**Rationale for the Four Constructs**

Each construct addresses a specific limitation in the existing theoretical landscape while drawing on a related body of scholarship. Epistemic partnership addresses the limits of connectivism's know-where account of learning by emphasizing the learner's work of evaluating and co-constructing knowledge with AI. The construct is grounded in distributed cognition and hybrid intelligence research. In response to cognitivism's difficulty in accounting for cognitive work shared across human and machine, distributed agency describes how responsibility, effort, and decision-making are allocated within human-AI interaction. Generative literacy extends constructivism by specifying the competencies learners need to remain actively engaged in meaning-making when AI can generate ready-made knowledge artifacts. Finally, adaptive metacognition addresses the limits of behaviorism's reliance on observable performance as evidence of learning by emphasizing learners' ongoing regulation of understanding, effort, and AI use. This construct draws on research on cognitive offloading and self-regulated learning.

The four constructs also cohere because they jointly answer the questions that any adequate account of learning must address. Alexander et al. (2009) argued that a defensible conceptualization of learning must specify what learning is, where it occurs, who the learner is, and how the process unfolds. Ertmer and Newby (2013) similarly compared learning theories by asking how learning occurs, which factors influence it, and how transfer takes place. Mapped onto these questions, epistemic partnership defines learning as knowledge co-construction in the relational space between learner and AI system. Distributed agency explains how cognitive work and responsibility are shared within that interaction. Generative literacy identifies the competencies learners need to participate in the partnership productively. Adaptive metacognition explains how learners monitor and regulate the process over time so that AI-supported co-construction contributes to durable individual understanding. These four constructs define the phenomenon, the structure of participation, the competencies required for enactment, and the regulatory process that sustains learning.

Importantly, the four constructs depend on one another. Epistemic partnership explains why learning with AI should be understood as a shared activity, but this partnership must be guided by clear decisions about who does what. Without distributed agency, learners may simply hand the task over to AI. Distributed agency, in turn, requires generative literacy. Learners cannot make sound decisions about AI use unless they know how to prompt, evaluate, revise, and integrate AI output. These literacy skills also need adaptive metacognition, because learners must decide when and why to use them during the learning process. Finally, metacognition matters most when there is an actual human-AI learning activity to monitor. In this way, the four constructs work together rather than separately. Empirical studies show what can happen when one part of this system is weak. In Darvishi et al.'s (2024) study, AI assistance improved performance in the short term, but the gains did not persist once the assistance was removed. Breakstone et al. (2021) pointed to a different risk: when learners lack evaluative skills, they may struggle to identify problems in fluent digital content. Fan et al. (2025) added that learners who did not monitor their own understanding showed signs of metacognitive laziness even when their task performance appeared adequate.

**Principles: Relations Among the Constructs**

The following four principles articulate the proposed relations among epistemic partnership, distributed agency, generative literacy, and adaptive metacognition. Figure 1 brings these relations together in an integrated conceptual model.

**Figure 1** *The Generativism Framework: Four Constructs and Four Principles*

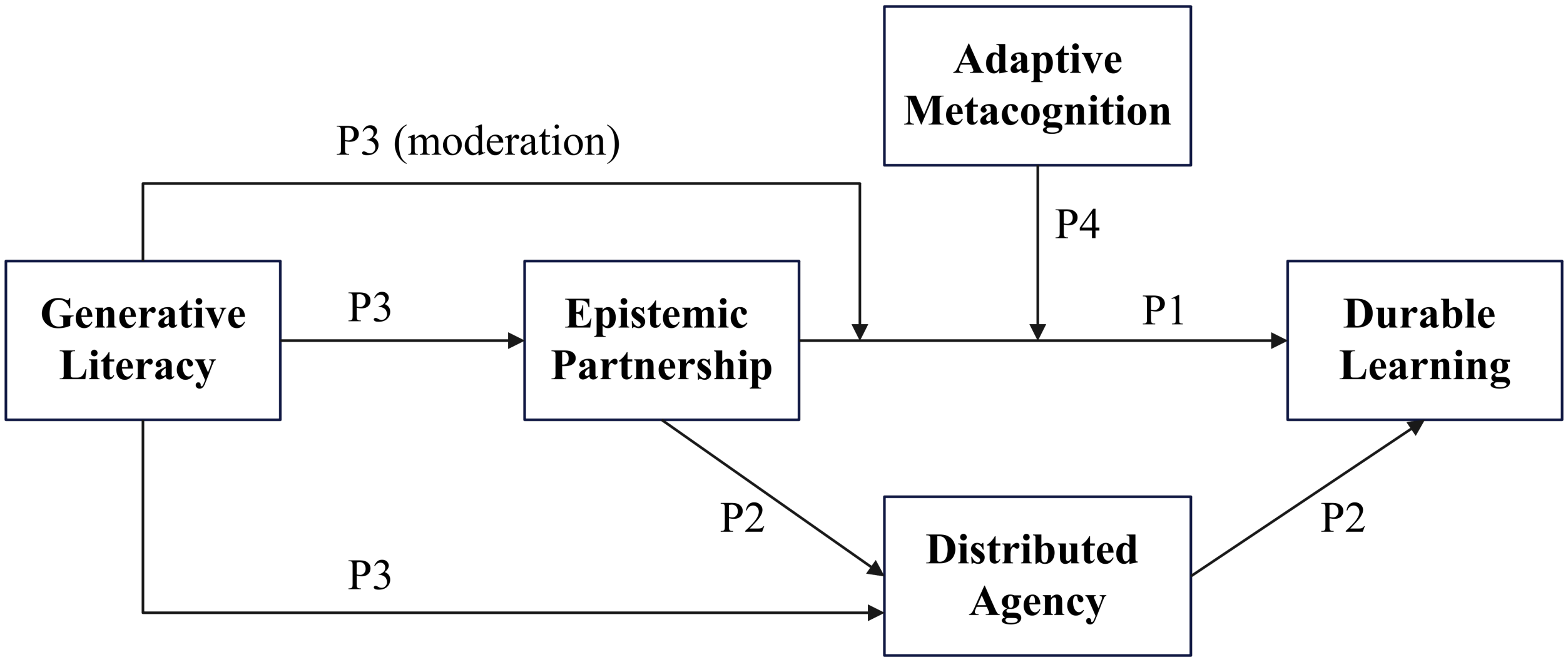

***Principle 1*** *Durable Learning Depends on the Quality of the Epistemic Partnership*

This principle holds that learning depends on the quality of the learner's epistemic engagement with a generative AI system (Wu et al., 2025). In a high-quality epistemic partnership, the learner treats AI-generated content as a provisional contribution to inquiry. The learner questions its assumptions, checks its claims, connects it with prior knowledge, and revises or rejects it when necessary. These activities require the learner to retain responsibility for interpretation and judgment, thereby preserving the cognitive work through which knowledge is constructed and later used without AI support. This principle is consistent with the ICAP framework, which predicts that learning increases as engagement moves from passive reception to constructive and interactive knowledge building (Chi & Wylie, 2014). Moreover, emerging empirical evidence suggests that the learning benefits of LLM interaction depend on learners' epistemic vigilance. Melumad and Yun (2025) reported the results of seven experiments in which 10,462 participants were randomly assigned to learn about a topic using either LLM-generated syntheses or conventional Google search results. They found that participants who used LLM summaries reported shallower knowledge and later produced advice that was sparser, less original, and less likely to be adopted. Melumad and Yun (2025) attributed this pattern to the reduced need for users to discover, compare, and synthesize information themselves. Ding et al. (2026) examined how university students interacted with ChatGPT when reviewing incorrect answers on an exam. They found that students who engaged in more active and constructive questioning showed greater improvement on later exam performance. Ruyang et al. (2026) focused more directly on *verification* in GenAI-supported writing. They studied two university EFL writing classes and examined how students responded to AI-generated feedback. Verification included behaviors such as asking the AI for explanations, refining prompts, and checking its suggestions against rubrics or teacher criteria. Students who engaged more frequently in these verification practices demonstrated deeper revisions and stronger transfer of learning. Similarly, Shen (2026) investigated 443 university students across Design and Business programs to understand how AI literacy, prompting, verification, and trust calibration related to coursework performance. They found that verification behavior and accurate trust calibration were associated with higher-quality coursework. In sum, these studies suggest that epistemic partnership quality is an important condition for durable learning with generative AI.

***Principle 2*** *Distributed Agency Mediates the Relation between Epistemic Partnership and Learning*

This principle proposes that epistemic partnership influences learning partly through the distribution of cognitive work between the learner and the AI system. Whereas epistemic partnership quality concerns how learners engage with AI-generated knowledge through questioning, verification, and sense-making, distributed agency concerns who performs particular cognitive operations and who retains control over consequential judgments. The constructs are therefore related but conceptually distinct. Distributed agency entails deliberate decisions about which cognitive operations to delegate to AI and which to retain (Hollan et al., 2000). These decisions matter because learning depends on the cognitive processes the learner actually performs. When AI handles explanation, comparison, or retrieval, immediate performance may improve, but the learner may have fewer opportunities to encode information or construct understanding (Fan et al., 2025). Choudhuri et al. (2026) provide evidence that the distribution of cognitive work may become increasingly unbalanced as trust in generative AI develops. Using survey data from 299 STEM students across five North American universities, the authors found that greater trust and routine use of generative AI were associated with lower self-reported reflection, need for understanding, and critical thinking. Prior experience with generative AI or higher education did not appear to protect students from this pattern. These findings do not directly measure distributed agency, but they indicate that routine reliance on AI may gradually reduce the cognitive operations learners retain for themselves. A well-developed sense of distributed agency is therefore important because it can make this allocation visible and support deliberate decisions about which forms of reasoning should remain under the learner's control.

***Principle 3** Generative Literacy Strengthens both Epistemic Partnership and Distributed Agency, and It Moderates the Effect of Epistemic Partnership on Learning*

Generative literacy gives learners the practical means to enact an epistemic partnership and to distribute agency deliberately. As defined earlier, generative literacy comprises four competencies: prompt competence, evaluative judgment, epistemic vigilance, and integrative synthesis. Prompt competence helps learners elicit explanations, alternatives, and feedback from AI systems. Evaluative judgment enables them to assess the accuracy, relevance, and coherence of these contributions, whereas epistemic vigilance sustains critical scrutiny even when the output appears fluent or plausible. Integrative synthesis enables learners to connect AI-generated material with prior knowledge and disciplinary understanding. These capacities improve epistemic partnership quality as they make the interaction more purposeful and more critically engaged. They also inform decisions about cognitive delegation. In this sense, we contend that generative literacy strengthens both epistemic partnership and distributed agency.

Moreover, generative literacy moderates the effect of epistemic partnership on learning. Mendoza et al. (2026) examined secondary students' use of ChatGPT and focused on whether feedback literacy shaped the relationship between students' engagement with AI-generated feedback and self-regulated learning. They found that feedback literacy significantly moderated this relationship. Students with stronger abilities to interpret, evaluate, and make sense of AI feedback benefited more from ChatGPT use, whereas those with lower feedback literacy showed weaker learning-related gains. Similarly, Ma and Guo (2026) investigated university students' GenAI-supported learning and examined AI literacy as a moderating factor in their learning model. Their findings showed that AI literacy altered the strength of relationships between GenAI-related factors and productive learning behavior.

***Principle 4*** *Adaptive Metacognition Moderates the Effect of Epistemic Partnership on Learning*

Adaptive metacognition regulates how learners engage with an epistemic partnership over time. Generative literacy provides the competencies needed to work productively with AI, whereas adaptive metacognition determines when and how those competencies are deployed. This proposition builds on self-regulated learning theories in which monitoring informs subsequent control and strategy adjustment (Winne & Hadwin, 1998). In AI-supported learning, however, regulation also involves managing the contribution of an external cognitive partner. Molenaar's (2022) Hybrid Human-AI Regulation framework supports this extension by describing how learners and AI systems adaptively share regulatory control. Liu et al. (2025) compared adaptive metacognitive scaffolding, planned metacognitive scaffolding, and a control condition in a generative AI-supported programming environment. They found that students receiving adaptive scaffolding demonstrated stronger performance on computational-thinking tasks and more complex problem-solving behavior than students in the other conditions. The adaptive scaffolding group also exhibited more recursive metacognitive regulation and lower mental effort.

More direct evidence for the proposed moderation comes from Bai (2026). In a quasi-experimental study of 108 university students, participants completed creative problem-solving tasks using generative AI, conventional search engines, or no technological support. Bai (2026) found that metacognitive regulation significantly interacted with the type of technological support in predicting aspects of problem-solving performance. Importantly, greater regulation did not uniformly produce better outcomes, suggesting that effective regulation depends on how it is calibrated to the task and the form of AI support. This finding supports treating metacognitive regulation as a boundary condition on the effects of human-AI interaction. It also reinforces the importance of the term *adaptive*: effective regulation requires adjustment to the particular epistemic demands of the interaction rather than simply more monitoring or control.

## Implications and Directions for Future Research

### Implications for Instructional Design

Generativism has direct implications for how learning environments are designed. Rather than banning AI or using it uncritically, the framework suggests designing learning activities that develop the competencies of epistemic partnership, agency awareness, generative literacy, and adaptive metacognition. A related design implication is the strategic use of *productive friction* (Ward et al., 2011). Because generative AI can reduce effort by delivering fluent, ready-made outputs, AI-integrated learning environments may need to deliberately preserve forms of difficulty that support understanding. Drawing on “desirable difficulties” (Bjork & Bjork, 2020) and “productive failure” (Kapur & Bielaczyc, 2012), designers can structure tasks so that AI prompts explanation, comparison, and justification. For example, instructional designers can require learners to attempt a solution before consulting AI, to diagnose and revise AI-generated work, or to reconcile competing AI-generated perspectives.

Several specific design strategies follow from the Generativist framework. First, learning activities can require learners to evaluate and improve AI-generated output rather than produce work from scratch. This approach leverages the production-comprehension inversion by using AI-generated products as starting points for critical analysis. Second, activities can incorporate structured prompting protocols that require learners to articulate their thinking before, during, and after engaging with AI. Research has shown that this design strategy promotes deeper cognitive engagement (Tomisu et al., 2025). Third, learning environments can make the distribution of cognitive labor between humans and AI visible and explicit. This can be achieved by requiring learners to document which aspects of a project were completed independently, which involved AI assistance, and how they evaluated and integrated AI contributions.

Moreover, designers should help learners make thoughtful decisions about when to rely on AI, question it, or retain responsibility for their own judgment, because learners are unlikely to distribute agency effectively by default. Logg et al. (2019) proposed algorithm appreciation, in which people give greater weight to advice when it comes from an algorithm than when it comes from a person. Dietvorst et al. (2015) documented algorithm aversion, in which people abandon an algorithm after observing even a small number of errors. These findings suggest that learners may over-delegate when AI appears confident or underuse AI when it makes visible mistakes.

**Implications for Learning**

From a generativist perspective, learning is increasingly expressed through the *process* of co-construction: how learners prompt, evaluate, revise, and integrate AI output, and how they decide what to delegate versus what to do themselves. Learners frequently begin with an AI-generated draft, solution, or explanation and then learn by interrogating it. This "production-comprehension inversion" offers a learning pathway that differs from those described by contemporary learning theories. In most instructional contexts, sophisticated production is assumed to be downstream from comprehension and therefore a reasonable proxy for learning. Nevertheless, a central learning phenomenon in the generative AI age is the weakening of the link between performance and understanding. When performance is no longer reliable evidence, the question becomes: what, then, should count as learning progress in human-AI work? First, learning under Generativism is evident when learners deepen their capacity for epistemic partnership. Specifically, learners treat AI output as a set of hypotheses to question, verify against sources or domain principles, and redirect when it becomes inaccurate or misaligned with the task. Moreover, learners show growth in distributed agency when they can deliberately allocate cognitive labor. They purposefully decide what to delegate to the system, what to retain for themselves, and how to preserve human authorship of key decisions. Additionally, learning is reflected in stronger generative literacy, including more precise prompting, sharper evaluative judgment, sustained epistemic vigilance, and the ability to effectively synthesize AI contributions. Lastly, learners develop adaptive metacognition as they become better at monitoring, controlling, and regulating their use of AI support for understanding and problem-solving. Ultimately, the goal is not AI-assisted output itself but the development of learners who can use generative AI to extend their thinking and address tasks while building durable understanding, transferable judgment, and responsibility for what they claim to know.

## Implications for Assessment

Assessment should differentiate between (a) contexts where AI collaboration is part of the target competence and should be assessed as such, and (b) contexts where independent performance is the intended outcome and AI use should be restricted. For contexts (a), assessments should examine how learners interact with AI, how they justify choices, and what they understand after the interaction. Table 2 summarizes assessment indicators aligned with the four generativist principles.

**Table 2**

*Indicators for Assessing Generativist Principles in Human-AI Learning*

| Constructs | Assessment indicator |
|---|---|
| Epistemic Partnership | The learner questions, verifies, redirects, and builds on AI output rather than accepting it as authoritative. |
| Distributed Agency | The learner can explain which parts of the task were delegated to AI, which were retained, and why that distribution was appropriate. |
| Generative Literacy | The learner demonstrates prompt revision, evaluation of accuracy and relevance, epistemic vigilance, and integration of AI output with prior knowledge. |
| Adaptive Metacognition | The learner monitors comprehension, detects confusion or overreliance, adjusts AI use, and reflects on how AI shaped their thinking. |

Evidence for the indicators can be gathered through process logs, annotated prompts, brief reflection statements, oral explanations or defenses, and comparisons of AI drafts with learner revisions. Recent higher education policy reviews show a shift from prohibition to guidance for integrating generative AI into assessment. Moorhouse et al. (2023), for example, argue that embracing GAI as part of the assessment process is beneficial because it reflects the realities of today's educational and job landscape. They conceptualize the instructor competence required for this shift as *generative artificial intelligence assessment literacy*. Separately, educators should verify what students can understand and do after AI-supported work. This can be assessed through short oral explanations or transfer tasks completed without AI, using rubrics that weigh judgment and reasoning in addition to product quality.

**Implications for Expertise Development**

Generative AI is transforming the nature of expertise in ways that existing learning theories do not address. Traditional conceptions of expertise emphasize the accumulation of domain knowledge and the development of automatized skills through sustained deliberate practice (Ericsson, 2004). Expert performance is characterized by rich, well-organized schemas, efficient pattern recognition, and reflective problem-solving within a domain. Cognitivist and constructivist theories serve these conceptions well by providing frameworks for understanding how schemas develop and how knowledge structures become more sophisticated over time. However, when generative AI can perform many tasks that traditionally defined expertise, the nature of expertise shifts. When factual information is readily available through AI, expertise depends less on possessing large amounts of information and more on knowing how to interpret, evaluate, and use that information well.

Generativism reconceptualizes expertise as the capacity for effective epistemic partnership. This means collaborating with AI in ways that leverage both human and AI strengths. Expertise entails knowing when to rely on AI, when to override it, and how to structure collaboration to produce outcomes that surpass what either partner could achieve alone. Under Generativism, the development of expertise involves not only acquiring domain knowledge and developing cognitive schemas but also cultivating the metacognitive, evaluative, and collaborative competencies that enable productive human-AI partnerships.

**Directions for Empirical Research**

Several lines of inquiry are particularly promising within a Generativism framework. First, studies examining how learners distribute cognitive effort between themselves and AI systems across different tasks and domains would provide evidence relevant to the principles of distributed agency and adaptive metacognition. In particular, this line of research calls for studies that reveal when and why learners transition from active engagement to passive acceptance of AI output, and what individual or contextual factors predict effective versus ineffective distributed agency. Moreover, intervention studies are needed to test whether deliberate instruction in generative literacy enhances effective human-AI collaboration. For instance, randomized controlled trials comparing learners who receive generative literacy instruction with those who do not would clarify its causal role in learning. Additionally, it would be fruitful to investigate how sustained use of generative AI affects the development of cognitive skills, metacognitive regulation, and domain expertise. Longitudinal designs could also track the development of adaptive metacognition over time, providing evidence about whether and how learners develop increasingly sophisticated strategies for managing human-AI collaboration. Furthermore, research examining how the principles of Generativism operate across diverse contexts would help establish the boundaries of the framework's applicability and identify necessary contextual adaptations. Lastly, design-based research investigating specific pedagogical implementations of Generativist principles would bridge the gap between theory and practice. Finally, research should test the architecture of the framework itself. Studies could examine whether the four principles are empirically separable, whether each contributes unique variance to learning outcomes in human-AI settings, and whether observed phenomena fall outside their joint scope.

**Limitations**

Several limitations of this work should be acknowledged. First, Generativism is a conceptual framework that has not been empirically tested. The constructs and principles proposed here are derived from a synthesis of existing evidence, but they require systematic empirical validation. Second, the framework does not fully address issues of access and equity, specifically the extent to which learners in different socioeconomic contexts have differential access to generative AI tools and the implications of this inequality for learning dynamics. Third, the rapid pace of AI development means that the specific technologies referenced in this article may evolve substantially, potentially altering the dynamics of human-AI learning in ways that the framework cannot anticipate.

## Conclusion

The four dominant learning theories of behaviorism, cognitivism, constructivism, and connectivism each captured important truths about learning in their respective eras. They remain valuable frameworks with continued relevance. However, the emergence of generative AI has introduced a set of learning dynamics that these theories do not fully explain. These include the capacity of non-human agents to generate knowledge artifacts, the redistribution of cognitive labor between human learners and AI systems, the paradox of enhanced performance coupled with reduced cognitive engagement, and the need for new forms of literacy oriented toward evaluating AI-generated content.

Generativism is offered as a preliminary response to these challenges. It is a theoretical framework organized around four constructs (epistemic partnership, distributed agency, generative literacy, and adaptive metacognition) and four principles specifying the relations among them. It draws on established scholarship in distributed cognition, extended mind, human-AI collaboration, AI literacy, cognitive offloading, and metacognition research. It does not claim to provide final answers but to articulate the question: How do we conceptualize learning when cognition is shared with a generative AI? Without an adequate theoretical framework, educators, designers, and policymakers lack the conceptual tools to make informed decisions about how generative AI should be integrated into learning. Generativism provides a lens for understanding what AI does to learning and what learning should look like in the AI age.

Like connectivism before it, which emerged as a response to the internet's transformation of knowledge and learning, Generativism emerges as a response to generative AI's transformation of what it means to produce, evaluate, and understand knowledge. Whether the generativism framework will prove sufficient to the challenge remains to be seen. What seems clear is that the learning challenges introduced by generative AI are substantive and that existing theories, while still valuable, are insufficient to address them independently. The invitation is to theorize, investigate, and refine our understanding of what it means to learn in an age when human cognition and artificial intelligence are becoming more closely integrated in learning activity.